# Valley filtering and electronic-optics using polycrystalline graphene

V. Hung Nguyen[1], S. Dechamps[1], P. Dollfus[2], and J.-C. Charlier[1]

[1]*Institute of Condensed Matter and Nanosciences, Université catholique de Louvain,
Chemin des étoiles 8, B-1348 Louvain-la-Neuve, Belgium*
[2]*Centre de Nanosciences et de Nanotechnologies, CNRS,
Univ. Paris Sud, Université Paris-Saclay, 91405 Orsay, France*

In this Letter, both the manipulation of valley-polarized currents and the optical-like behaviors of Dirac fermions are theoretically explored in polycrystalline graphene. When strain is applied, the misorientation between two graphene domains separated by a grain boundary can result in a mismatch of their electronic structures. Such a discrepancy manifests itself in a strong breaking of the inversion symmetry, leading to perfect valley polarization in a wide range of transmission directions. In addition, these graphene domains act as different media for electron waves, offering the possibility to modulate and obtain negative refraction indexes.

Due to its peculiar electronic structure and outstanding properties, graphene has become a material of choice for many fundamental researches and promising applications [1, 2]. The attractiveness of graphene lies basically in its massless Dirac fermions, high carrier mobility, small spin-orbit coupling, optical transparency, superior mechanical properties, and so on. Several fascinating phenomena (e.g., Klein tunneling, anomalous quantum Hall effect, Berry's phase manifestation, etc.) have been explored [1]. Graphene also turns out to be very promising for integration into a variety of electrical, spintronic, optical applications, flexible electronics, and so on [2]. Especially, two unconventional concepts, *valleytronics* and *electronic-optics*, have been explored and rapidly attracted a great amount of attention from various scientific communities, e.g., see [3] and references therein.

Valleytronics [4] lies in exploiting the feature that charge carriers flow through graphene as a wave populating the $K$ and $K'$ valleys [1] in its Brillouin zone, with each valley being characterized by a distinct momentum and a valley index [5, 6]. In analogy with spintronics, manipulating carriers in these two valleys can be used to encode data, i.e., to represent the zeroes and ones in digital computing. To date, many strategies [5–9] have been proposed to break the valley degeneracy for creating and detecting the valley polarization in graphene. They mainly rely on the valley filtering effects and/or the generation of spatially-separated valley-resolved currents in graphene nanostructures. Most remarkably, recent experimental advances have successfully demonstrated the electrical generation and control of pure valley current in graphene systems where the inversion symmetry is broken by a gap-opening perturbation [10, 11].

The high mobility of carriers in graphene allows for ballistic transport over micrometer length scales even at room temperature [12]. Hence, electrons can flow in straight-line trajectories and their wave nature can manifest in a variety of interference and diffraction effects [13], in analogy with light rays in optical media. This makes graphene an ideal platform for demonstrating electronic-optics and hence for developing novel quantum devices [14–19]. In this respect, the advantages of graphene also come from peculiar properties as low-energy linear dispersion with electron-hole symmetry, gapless character and excellent gate controllability [17–19]. Due to the specific electronic dispersion, the group velocity of electrons (resp. holes) in graphene is parallel (resp. antiparallel) to their momentum. This results in negative refraction, a striking feature of electronic-optics, when carriers transmit across graphene p–n junctions [14, 19]. The gapless character makes graphene nanostructures highly transparent [15] while the gate controllability offers possibilities of electrically modulating the refraction index in electronic-optics components [16–19].

However, for practical applications, graphene samples synthesized at large scale (e.g., by CVD) are always found to be polycrystalline in nature [20–22], i.e., composed of different single-crystal grains separated by grain boundaries (GB). This defective nature strongly affects the intrinsic properties of graphene [21, 22]. Nevertheless, polycrystalline graphene can also offer opportunities of tailoring the electronic properties of domains so as to achieve desirable properties of the global system [21–30]. In this respect, several experiments have been conducted to characterize and control individual grains and grain boundaries, allowing for both intra- and intergrain transport measurements [23–26]. In the same direction, it has been predicted that besides being affected by the defect scattering [23], the transport through polycrystalline graphene strongly depends on the symmetry properties of its domains, which could be exploited to tune and achieve a metal-semiconductor transition [29, 30]. In particular, if single-crystal domains are arranged in different orientations, a mismatch of their electronic structures can occur and hence a finite energy gap of conductance (transport gap) opens [29]. Actually, this gap is not an electronic bandgap of graphene, i.e., it is achieved though the single-crystal domains remain semimetallic. The misorientation between these domains also offers the opportunity to modulate the transport gap by strain engineering [30].

In this Letter, a new and practical scheme is pro-posed to manipulate highly valley-polarized currents and optical-like behaviors of charge carriers in graphene, which are essential ingredients for valleytronics and electronic optics applications. The approach lies in the use of polycrystalline graphene systems containing two misoriented domains separated by a GB, that is composed of a periodic array of dislocations (see the examples in Fig.1.a-b). In principle, these two domains exhibit different electronic structures, especially, when the system is strained. Such a discrepancy is expected to manifest itself in a strong breaking of the inversion symmetry of the system and hence a high valley polarization can be achieved. Additionally, in analogy to optical systems, these domains act as different media for electron waves, leading to optical-like behaviors of charge transport through the system. It is worth noting that this general approach can be extended to several polycrystalline systems of other materials that, similarly, contain different crystalline domains, e.g., see in [31, 32].

Our investigation was conducted using atomistic simulations described in [30]. Considered graphene systems contain two commensurable domains separated by a periodic GB along the y-axis (see two examples in Fig.1.a-b). The atomistic geometry of the system was relaxed using molecular dynamics to minimize its energy determined from optimized Tersoff potentials [33]. A uniaxial strain is applied with magnitude $\sigma$ and direction $\theta$ (see the inset of Fig.1.a-b). The transport across the GB was computed using Green's function method to solve the tight-binding model that has been empirically constructed to include the effects of strain [34, 35].

Generally, charge transport through graphene-GB systems can be easily modulated by strain. Indeed, when strain is applied, Dirac cones (or valleys) $D$ and $D'$ of graphene are displaced and no longer located at the $K$ and $K'$ points [34]. Most importantly, misoriented domains show different responses to strain, leading to a separation of their Dirac cones in the $k$-space. Without any additional scattering, the charge transport satisfies the conservation of momentum $k_y$. The misalignment of Dirac cones hence results in a finite transport gap, which has been shown to be a function of strain ($\sigma$ and $\theta$) and lattice symmetry [30]. Let us now consider a de Broglie wave of electron approaching the GB from the left domain, which is characterized by velocity $\mathbf{v_1} = v_1(\cos\phi_1, \sin\phi_1)$ and momentum $\mathbf{k_1} = \mathbf{D_1} + \mathbf{q_1}$ where $\mathbf{q_1} = q_1(\cos\phi_1, \sin\phi_1)$ and $\phi_1$ is the incident angle. At the GB, this wave is partly reflected to the state with $\mathbf{v_1^r} = v_1(-\cos\phi_1, \sin\phi_1)$ and partly transmitted to the state with $\mathbf{v_2} = v_2(\cos\phi_2, \sin\phi_2)$ and $\mathbf{k_2} = \mathbf{D_2} + \mathbf{q_2}$ [$\mathbf{q_2} = q_2(\cos\phi_2, \sin\phi_2)$] in the right domain (see Fig.1.c). Due to the $k_y$-conservation, the equation $q_1\sin\phi_1 - q_2\sin\phi_2 = D_{2y} - D_{1y}$ is satisfied. A similar relationship is obtained for valley $D'$ with $D'_{2y} - D'_{1y} = -(D_{2y} - D_{1y})$. Under small strain, the difference of energy dispersion around the Dirac cones between the two graphene domains is negligible, leading to $q_1 \cong q_2 = q$. This work focuses on only small strains that are much lower than the bandgap threshold $\sigma_{gt} \simeq 23\%$ [34]. The above equation for $\phi_{1,2}$ can be hence rewritten as

$$\sin\phi_1 - \sin\phi_2 = \eta_v \alpha(q), \quad (1)$$

where $\alpha(q) = (D_{2y} - D_{1y})/q$ and $\eta_v = \pm 1$ for valleys $D$ and $D'$, respectively. Furthermore, the group velocity and momentum are parallel (resp. anti-parallel) for electrons (resp. holes). Equation (1) has to be modified, i.e., changing the sign of rhs term, for holes. In addition, in graphene under small strain, the energy dispersion around Dirac points is still linear and can be described by $\varepsilon = \pm\hbar v_F q$ with $v_F \simeq 1 \times 10^6$ m/s [1]. Hence, equation (1) can apply to both electrons and holes by using $\alpha(q) = \hbar v_F (D_{2y} - D_{1y})/\varepsilon$. Note additionally that the similar separation of Dirac cones and hence the same refraction rule can be achieved in incommensurable systems without strain [29]. In such a case, the strain can be also used to modulate these transport properties of the system [30, 35].

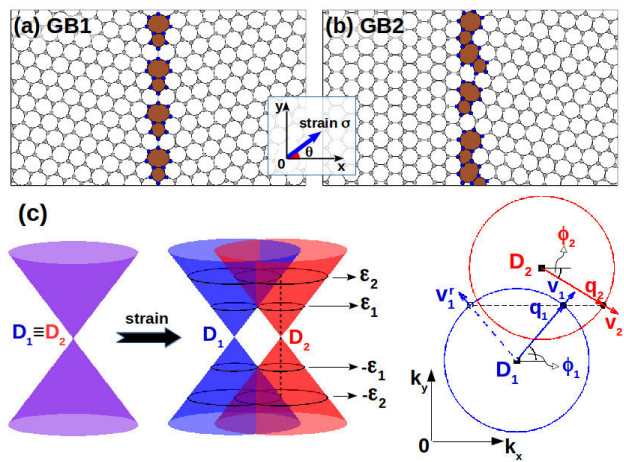

FIG. 1. Atomistic structure of graphene grain boundary systems $(2,1)|(1,2)$ (a) and $(0,7)|(3,5)$ (b). Charge carriers transmitted across the boundary under a uniaxial strain of magnitude $\sigma$ and direction $\theta$ are considered. (c) Diagrams illustrating the strain effects on the band structure of graphene domains (left) and the momentum conservation rule (right).

Clearly, equation (1) implies that transmission in the two valleys is modulated differently by strain. Simultaneously, it manifests itself in a new refraction rule, compared to the ordinary ones, i.e. $\sin\phi_2 = \pm\frac{q_1}{q_2}\sin\phi_1$, in graphene doped structures [14, 15]. Equation (1) is thus the key-element of the current work, which can explain high valley polarization and optical-like behaviors of charge carriers presented below.

***Valley filtering effects.*** In Fig.2, transmission functions $\mathcal{T}_{D,D'}$ in the two valleys and valley polarization $P_{val} = (\mathcal{T}_D - \mathcal{T}_{D'})/(\mathcal{T}_D + \mathcal{T}_{D'})$ in the two graphene systems GB1 and GB2 (see Fig.1.a-b) are displayed as a function of incident angle $\phi_1$ for different applied strains. Without strain, similarly to the case of graphene systems containing a line defect [7], scattering at GBs is shown to modify differently the functions $\mathcal{T}_{D,D'}$, leading to a finite $P_{val}$. However, a high $P_{val}$ is obtained only for large incident angles. When strain is applied, gaps of $\mathcal{T}_{D,D'}$ open in different ranges of $\phi_1$ and hence $P_{val}$ can be strongly enhanced. Indeed, in both systems GB1 and GB2, a perfect $P_{val}$ can be achieved in wide ranges of $\phi_1$ when applying a reasonably large strain, i.e., as shown for $\sigma = 2 \div 3\%$ in Fig.2. Additionally, $P_{val}$ can be also modulated by tuning carrier energy, as discussed below.

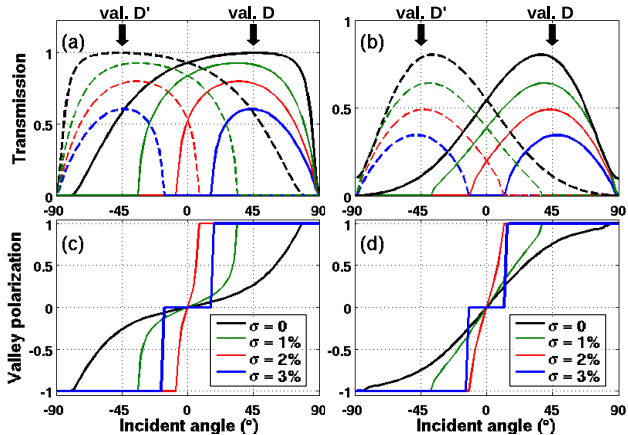

FIG. 2. (a,b) Transmission probability in the two valleys and (c,d) valley polarization as a function of incident angle for energy $\varepsilon = 0.3$ eV and different strains. (a,c) and (b,d) present the data obtained in the systems GB1 (for strain direction $\theta = 45°$) and GB2 (for $\theta = 22°$) in Fig.1.a-b, respectively.

To clarify these properties, equation (1) is used to distinguish different transport regimes as illustrated in Fig.1.c. The $k_y$-conservation implies that the transmission is non-zero only if the solution of equation (1) exists. The angle $\phi_1$ has thus to satisfy $\max(\eta_v \alpha(q) - 1, -1) \le \sin\phi_1 \le \min(\eta_v \alpha(q) + 1, 1)$. In the low energy regime $|\varepsilon| < \varepsilon_1$ (i.e., $|\alpha(q)| > 2$) with $\varepsilon_1 = \hbar v_F |D_{2y} - D_{1y}|/2$, no angle $\phi_1$ satisfies the above condition and thus the system is totally reflective, corresponding to a transport gap. In the second regime $\varepsilon_1 \le |\varepsilon| < \varepsilon_2$ (i.e., $1 < |\alpha(q)| \le 2$) with $\varepsilon_2 = \hbar v_F |D_{2y} - D_{1y}|$, the transmission is found to be non-zero in finite ranges of $\phi_1$. In particular, to obtain non-zero transmission of electrons, $\phi_1$ has to satisfy $\alpha(q) - 1 \le \sin\phi_1 \le 1$ and $-1 \le \sin\phi_1 \le 1 - \alpha(q)$ for valleys $D$ and $D'$, respectively. These two ranges are fully separated, i.e., $\phi_1 > 0$ and $\phi_1 < 0$, respectively. In such ranges, the transmission is allowed for only one valley, leading to perfect $P_{val}$ as observed for $\sigma = 3\%$ in Fig.2. The third regime is $|\varepsilon| \ge \varepsilon_2$, i.e., $|\alpha(q)| \le 1$. In this regime, the ranges of $\phi_1$ allowing for finite $\mathcal{T}_{D,D'}$ are larger than those observed in the second one and can overlap. Similarly as above, outside the overlapped range of $\phi_1$, a perfect $P_{val}$ is obtained. Otherwise, $P_{val}$ has smaller values inside the overlapped range, as shown for $\sigma = 1\%$ and $2\%$ in Fig.2.

**Optical like behaviors of charge carriers.** As mentioned earlier, equation (1) also implies that electrons transmitted across the system have optical-like behaviors with new refraction rules. The relationship between $\phi_{1,2}$ extracted from the data in Fig.2 is displayed in Fig.3.a-b. Note that without strain and/or in systems containing domains of same orientation [7], $\alpha(q) = 0$ and hence $\phi_2 = \phi_1$. Otherwise, the refraction index can be easily modulated by strain and/or by varying carrier energy, i.e., $\sin\phi_2 = \sin\phi_1 - \eta_v \alpha(q)$. Indeed, for small strains (e.g., $\sigma = 1\%$ and $2\%$ here), both positive and negative refraction indexes are observed. Interestingly, when strain is large enough (i.e., $\sigma = 3\%$), the refraction index is negative in the full $\phi_{1,2}$-ranges of finite transmission. Actually, the former case (small strains) corresponds to the energy regime $|\varepsilon| \ge \varepsilon_2$ while the latter corresponds to $\varepsilon_1 \le |\varepsilon| < \varepsilon_2$ discussed above. Moreover, due to the equivalence between $\phi_1$ and $-\phi_2$ in equation (1), the solutions of $\phi_{1,2}$ are symmetric with respect to $\phi = 0$.

Other novel properties are predicted as presented in Fig.3.c. First, even being non-zero in a finite range of $\phi_{1,2}$, $\mathcal{T}_{D,D'}$ are numerically shown to be high only for directions around a specific angle satisfying $\phi_2 \cong -\phi_1 = -\phi_h$ where $\phi_h = \eta_v \arcsin(\alpha(q)/2)$. This angle $\phi_h$ can be tuned by changing carrier energy and/or strain, e.g., $\phi_h \simeq \pm 40°$ for valleys $D$ and $D'$ in the case of $\varepsilon = 0.3$ eV and $\sigma = 3\%$ presented here. This property can be an important ingredient for controlling directional currents and highly focused beams. Note that in graphene p–n junctions, only the beams around $\phi = 0$ are highly transmitted while the transmission is low for large angles, especially, when the transition length between highly doped regions is large [15]. This essentially limits the performance of corresponding electronic-optics devices, especially, at high temperature [19]. Moreover, when injected into the graphene-GB system, carriers in one valley are totally reflected while partly for other valley, implying that both transmitted and reflected beams are highly valley-polarized. In p–n junctions, these two beams can be also achieved but are valley-unpolarized [15, 18].

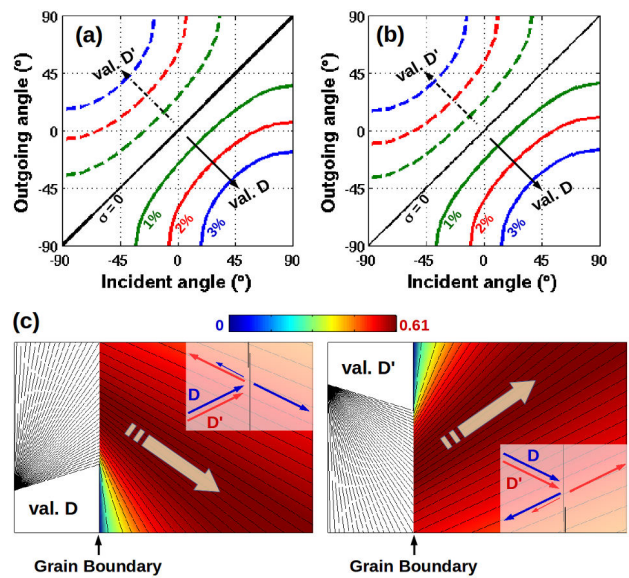

FIG. 3. Strain-induced modulation of refraction index in the GB1 (a) and GB2 (b) systems extracted from the data in Fig.2. (c) Diagrams illustrating the properties of electron beams with negative refraction (particularly, for $\sigma = 3\%$ in (a)) and color-maps showing the transmission function for different outgoing angles.

***Directionally separated currents.*** For practical perspectives, it is necessary to analyze the electrical currents, which are always measured at finite temperature, contributed by carriers with different energies and, importantly, which flow in different directions [19]. This analysis can be performed by rewriting the standard Landauer formula for the left-to-right current and then extracting these directional components. In particular, the conductance can be computed as $G = \int_{-\pi/2}^{\pi/2}[\mathcal{G}_D(\phi_2) + \mathcal{G}_{D'}(\phi_2)]\cos\phi_2 d\phi_2$ with

$$\mathcal{G}_{D,D'}(\phi_2) = \frac{e^2}{\pi h}\frac{W}{\hbar v_F}\int \mathcal{T}_{D,D'}(\epsilon,\phi_2)[-\frac{\partial f(\epsilon)}{\partial \epsilon}]|\epsilon|d\epsilon \quad (2)$$

where, $W$ denotes the width of graphene sheet and $f(\epsilon)$ is the Fermi distribution function. Thus, the left-to-right current is the sum of contributions of all components flowing in different directions. In practice, these directional currents can be experimentally measured using multiple directional leads [10, 11, 18, 19, 23–26], as diagrammatically simplified in Fig.4.e. Actually, the directional currents partly contribute to the left-to-right one measured at the right lead and partly reach the leads $T$ and $B$, depending on their intensity and direction.

In Fig.4.a, conductances $\mathcal{G}_{D,D'}(\phi_2)$ computed using equation (2) and valley polarization $P_{val} = (\mathcal{G}_D - \mathcal{G}_{D'})/(\mathcal{G}_D + \mathcal{G}_{D'})$ at different temperatures are presented. Actually, directionally separated currents with nearly perfect $P_{val}$ and high intensity around the directions $\phi_2 = -\phi_h \simeq \mp 40°$ (for valleys $D$ and $D'$, respectively) are achieved even at 300 $K$. Complete pictures showing the dependence of $\mathcal{G}_{D,D'}$ and $P_{val}$ on the Fermi level $E_F$ are also displayed in Fig.4.b-d. As discussed above, when varying $E_F$ (e.g., by tuning a back gate voltage [19]), the direction $\phi_h$ of high intensity current can be easily modulated as it moves to small angles when increasing $E_F$ and vice versa. Most importantly, the well-separated currents and high $P_{val}$ can be observed at room temperature and in a wide range of $E_F$ as shown for $|E_F| \leq 0.4$ eV here.

Remarkably, when changing from electrons to holes, $\mathcal{G}_{D,D'}$ reverse their direction (see Fig.4.b-c). Accordingly, $P_{val}$ reverses its sign (see Fig.4.d). This feature is essentially a consequence of the fact that the group velocity of electrons (resp. holes) is parallel (resp. anti-parallel) to the momentum. This leads to a change in the sign of $\phi_{1,2}$ as described in equation (1) when applied to hole transport. With the gate-tunability of $E_F$ [17–19], this property offers an excellent possibility of gate controlling the valley polarized currents in considered devices.

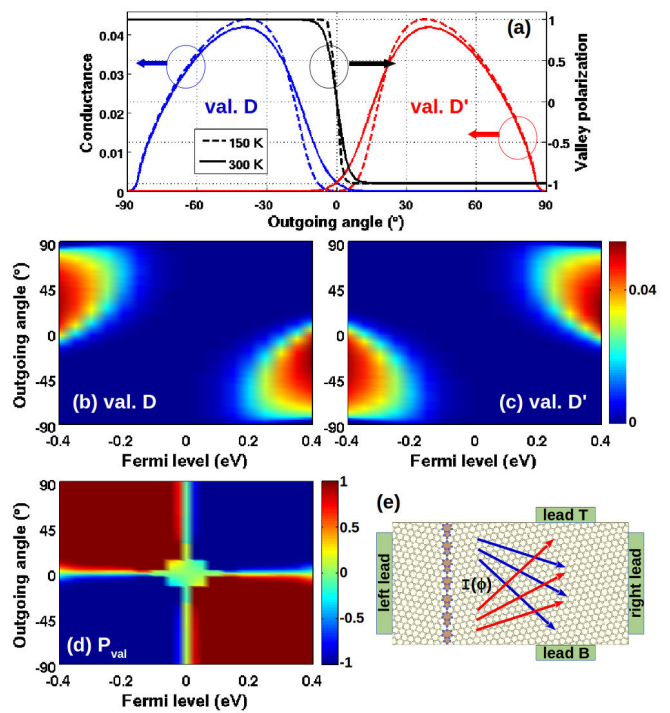

FIG. 4. (a) Conductance $\mathcal{G}_{D,D'}$ (left axis) in unit of $\frac{2e^2 W}{3\pi h a_0}$ ($a_0 = 1.42$ Å) and valley polarization $P_{val}$ (right axis) as a function of outgoing angle $\phi_2$ for Fermi level $E_F = 0.3$ eV and at different temperatures. $(E_F, \phi_2)$-maps of $\mathcal{G}_{D,D'}$ (b,c) and $P_{val}$ (d) at 300 $K$. All data were obtained for $\sigma = 3\%$ in the GB1 system. (e) Simple schematic of multiple electrode device for measuring the directional currents.

Finally, though the effects of scattering at grain boundaries on the transmission function are strongly dependent on their atomic structure (e.g., see Fig.2), our calculations show that the properties discussed above should be common to all periodic grain boundaries.

In conclusion, polycrystalline graphene, when subjected to strain, has been demonstrated to be a very promising candidate for manipulating highly valley-polarized currents and optical-like behaviors of massless Dirac fermions with novel refraction rules. Interestingly, the predicted properties are observable at room temperature, in a wide range of incident angle and energy, and gate-controllable. Although limited to the periodic grain boundaries, the study presents several practical perspectives. Generally, the disorder at grain boundaries could affect the $k_y$-conservation and hence induce a leakage current within the transmission gaps. However, some efforts to achieve periodic grain boundaries at large scale have been experimentally realized [23, 27, 28]. In particular, periodic grain boundaries as long as a few ten nanometers can be obtained after thermal reconstruction of aperiodic ones [28]. Otherwise, a weak disorder is expected to not strongly affect the transport properties observed in periodic systems, e.g., see in [29, 36]. Following the basic principles, momentum conservation and mismatch of electronic properties between different crystalline domains, our general approach can be extended to other polycrystalline systems of different materials including 2D layered, thin films and 3D materials [31, 32].

V.H.N. and J.-C.C. acknowledge financial support from the F.R.S.-FNRS of Belgium through the research project (N° T.1077.15), from the Commmunauté Wallonie-Bruxelles through the ARC on Graphene Nanoelectromechanics (N° 11/16-037) and from the European Graphene Flagship (N° 604391).

# Supplemental material for
# "Valley filtering and electronic-optics using polycrystalline graphene"

V. Hung Nguyen[1], S. Dechamps[1], P. Dollfus[2], and J.-C. Charlier[1]

[1]Institute of Condensed Matter and Nanosciences, Université catholique de Louvain,
Chemin des étoiles 8, B-1348 Louvain-la-Neuve, Belgium
[2]Centre de Nanosciences et de Nanotechnologies, CNRS,
Univ. Paris Sud, Université Paris-Saclay, 91405 Orsay, France

## 1. Tight binding approach versus first-principles calculations

In this section, we present a comparison between the tight binding approach employed in our study and first-principles calculations.

Although first principles computation methods have been known as one of the most accurate techniques for the description of the electronic properties of materials, their computational cost is quite expensive and they can not be conveniently performed to investigate systematically the quantum transport in complicated hetero-structures. In this regard, the tight binding (TB) approach is another alternative and good choice to study polycrystalline graphene systems and, especially, to take into account the effects of any arbitrary strain. In this study, we hence employed the quantum transport calculations within a TB Hamiltonian described in [1]. This TB model is empirically parametrized as in [2] to include the effects of strain.

To confirm the validity of our computation methods, the comparison between this TB approach and calculations based on the density functional theories (DFT) was made in simple cases for both pristine graphene and graphene grain boundary systems.

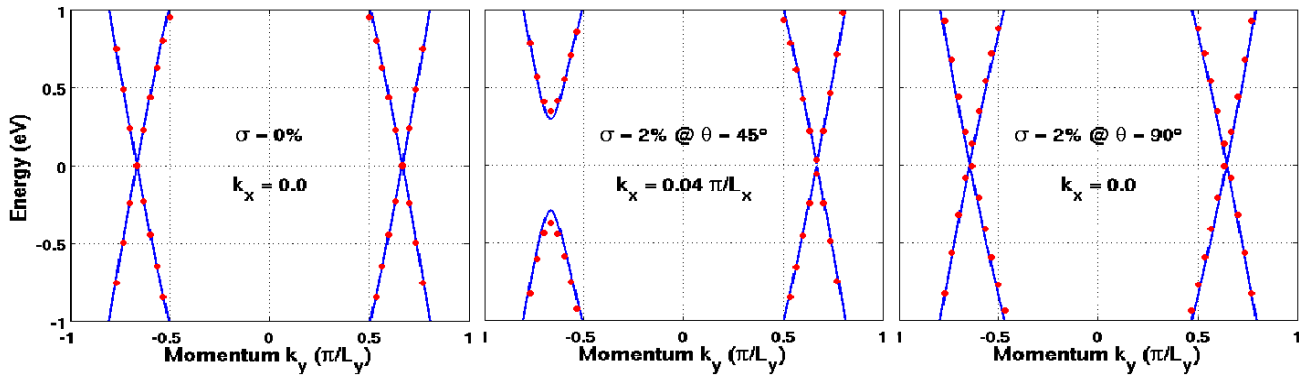

**Fig. S1**: Bandstructure of pristine graphene with different applied strains:
comparison between TB (solid lines) and DFT (closed circles) calculations.

First, since the considered systems contain two semi-infinite (i.e., large) pristine graphene domains separated by a single grain boundary, the accurate description of the electronic properties of pristine graphene is necessary. In the low energy range around the Dirac point, it is well known that the TB approach can compute very accurately the electronic properties of pristine monolayer graphene [3], compared to the DFT calculations. Indeed, this is confirmed in Fig.S1 that both methods are in very good agreement for describing the low energy bandstructure of graphene in both cases without/with strain. Here, the DFT band structure of pristine graphene was computed using the ABINIT code [4].

In graphene grain boundary systems, the transport properties are actually governed by two main factors: the electronic properties of two pristine graphene domains surrounding the grain boundary and scattering at the grain boundary. As explained in our main paper, the electronic properties of these graphene domains play decisive roles on the transport gap opening and their mismatch is the principal origin of high valley filtering and electronic optics effects explored. Regarding this point, as demonstrated above, the TB method is shown to be a very good approach, compared to the DFT calculations.

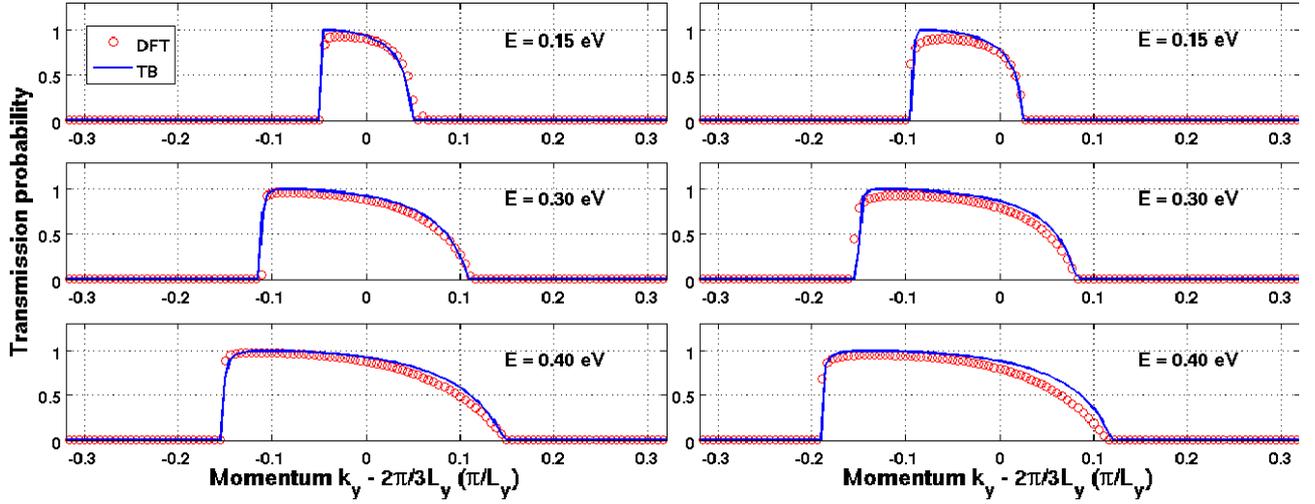

**Fig. S2**: Transmission probability in the system GB1 without (left panels) and with strain $\sigma = 2\%$ (right panels): comparison between TB and DFT calculations.

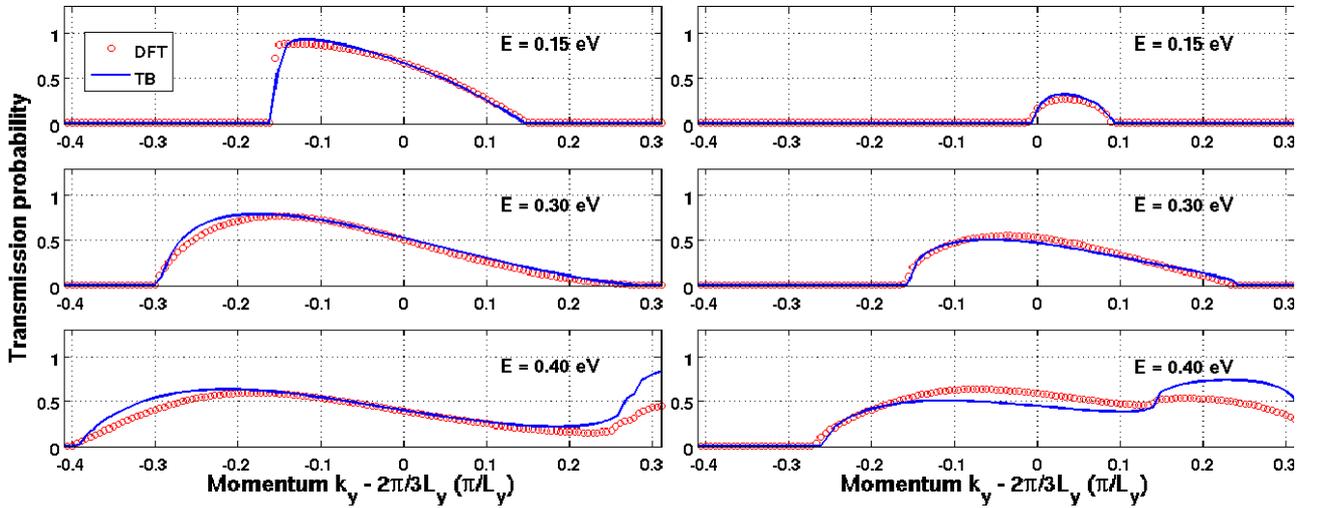

**Fig. S3**: Transmission probability in the system GB2 without (left panels) and with strain $\sigma = 2\%$ (right panels): comparison between TB and DFT calculations.

The grain boundary plays another important role, i.e., inducing defect scattering that can affect the transport properties of the system beyond the gap regions. Generally, its effects degrade the transmission probability, compared to the pristine graphene case. In Figs. S2 and S3, we display the transmission probability for obtained for two systems GB1 and GB2, respectively, as a function of momentum $k_y$ for different carrier energies, without and with strain, showing the comparison between TB and DFT calculations. The DFT transport calculations were performed using the OpenMX code [5]. Indeed, in the gap regions where the transmission function is zero, the two methods show a perfect agreement. Beyond the gaps, their agreement is also very good in the energy range (low energies) investigated in our study, thus confirming the validity of the TB calculations for the prediction of electronic transport properties presented in this Letter.

However, we also notice that in the case of more highly defective graphene systems, i.e., containing many grain boundaries with high defect density (e.g., as in [6]), defect scattering or effects of localized states at the grain boundary should be much stronger and hence this simple TB model may not be appropriate to compute accurately the electronic transport. In such a case, a more complicatedly parametrized TB model should be constructed, e.g., as performed in [7].

## 2. Valley filtering and electronic optics effects in incommensurable systems

Graphene grain boundary systems can be classified into two main classes [8]. The class I contains the systems where two graphene domains exhibit similar electronic structures at low energies, i.e., their Dirac cones are located at the same k point [at $(0,\pm 2\pi/3L_y)$ or $(\pm 2\pi/3L_x,0)$]. In the systems of class II, two graphene domains exhibit different electronic structures, i.e., Dirac cones of one domain are located at $\mathbf{k} = (0,\pm 2\pi/3L_y)$ while they are located at $\mathbf{k} = (\pm 2\pi/3L_x,0)$ for the other domain. While the commensurable systems are always belong to the class I, the incommensurable ones can be either in the class I or class II. The incommensurable systems of class I have similar properties as presented in our main paper but the systems of class II can exhibit some different properties.

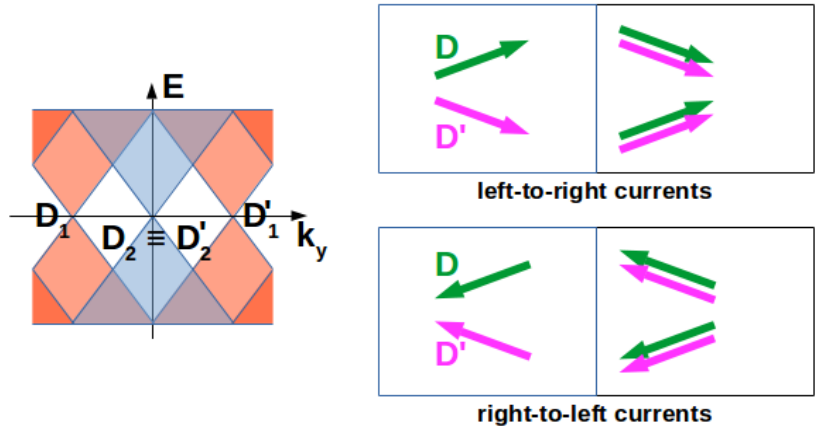

**Fig. S4**: Bandstructure profile (left) of two graphene domains in the case of
class II without strain. Diagrams illustrating the transport properties through the system.
Note that the reflected beams are not presented.

Since our reported results are essentially based on the conservation of momentum $k_y$, the separation of Dirac cones of two graphene domains in the $k_y$-axis is the key point. In the systems of class I, such the separation can be generated by strain effects [1]. For the class II, this separation (see in Fig. S4) are present even without strain, as discussed in [8]. Because of this property, valley filtering and electronic optics effects are still observed in these systems but the situation is a little bit different from what observed for the systems of class I.

For the class II in the unstrained case, as illustrated in diagrams of Fig. S4, charge carriers in each valley of domain 1 can transmit to both valleys of domain 2. Hence, outgoing currents from left to right are valley-unpolarized. On the contrary, they are highly valley-polarized for currents from right to left. In both cases, electronic optics behaviors are still observed.

When strain is applied, these properties can be modified. Due to strain effects, the two valleys

$D_2$ and $D_2'$ of right domain can be separated in the $k_y$ axis so that one of them approaches the valley $D_1$ of left domain and the other one is closer to the valley $D_1'$ (see in Fig.S5). Because of this feature, the same properties as presented in our main paper can be observed in these systems in the energy ranges $E \in [\varepsilon_1,\varepsilon_2]$ and $[-\varepsilon_2,-\varepsilon_1]$.

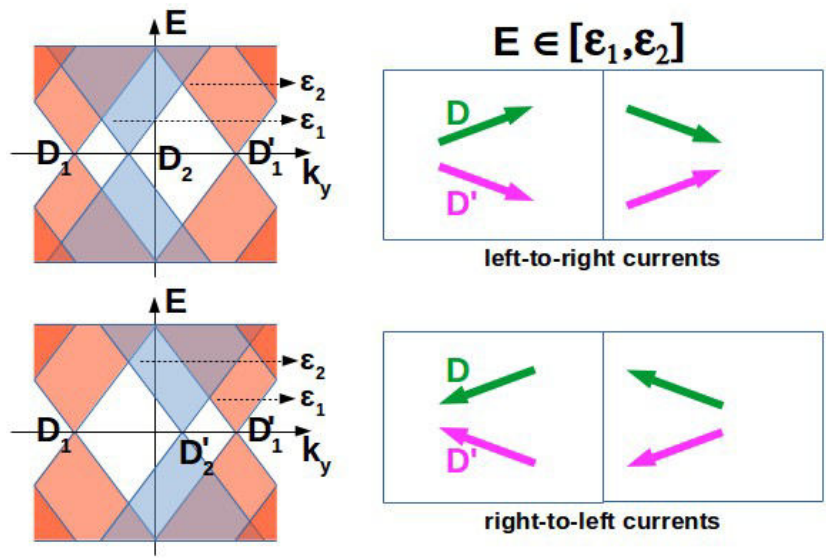

**Fig. S5**: Bandstructure profile (left) of two graphene domains in the case of class II with strain. Diagrams illustrating the transport properties through the system. Note that the reflected beams are not presented.